\documentclass[conference]{IEEEtran}
\IEEEoverridecommandlockouts
\usepackage{cite}
\usepackage{amsmath,amssymb,amsfonts}
\usepackage{algorithmic}
\usepackage{graphicx}
\usepackage{textcomp}
\usepackage{xcolor}
\usepackage{amsmath}
\usepackage{makecell}
\usepackage{mathrsfs}
\usepackage[ruled,vlined]{algorithm2e}
\usepackage{booktabs,multirow,multicol,color}
\usepackage{threeparttable}
\def\BibTeX{{\rm B\kern-.05em{\sc i\kern-.025em b}\kern-.08em
    T\kern-.1667em\lower.7ex\hbox{E}\kern-.125emX}}
\begin{document}
\bibliographystyle{plain}

\title{URGENT-PK: Perceptually-Aligned Ranking Model Designed for Speech Enhancement Competition}

\author{
\IEEEauthorblockN{
$^{1}$Jiahe Wang,
$^{1}$Chenda Li,
$^1$Wei Wang,
$^1$Wangyou Zhang,
$^2$Samuele Cornell,
$^3$Marvin Sach,
$^4$Robin Scheibler, \\
$^5$Kohei Saijo, 
$^3$Yihui Fu, 
$^6$Zhaoheng Ni, 
$^4$Anurag Kumar,
$^3$Tim Fingscheidt,
$^2$Shinji Watanabe,
$^1$Yanmin Qian
}
\IEEEauthorblockA{ 
\textit{$^1$Shanghai Jiao Tong University, China
$^2$Carnegie Mellon University, USA}
}
\IEEEauthorblockA{
\textit{
$^3$Technische Universität Braunschweig, Germany
$^4$Google DeepMind, Japan
$^5$Waseda University, Japan
$^6$Meta, USA}
}
}

\maketitle

\begin{abstract}
The Mean Opinion Score (MOS) is fundamental to speech quality assessment. However, its acquisition requires significant human annotation. Although deep neural network approaches, such as DNSMOS and UTMOS, have been developed to predict MOS to avoid this issue, they often suffer from insufficient training data. Recognizing that the comparison of speech enhancement (SE) systems prioritizes a reliable system comparison over absolute scores, we propose URGENT-PK, a novel ranking approach leveraging pairwise comparisons. URGENT-PK takes homologous enhanced speech pairs as input to predict relative quality rankings. This pairwise paradigm efficiently utilizes limited training data, as all pairwise permutations of multiple systems constitute a training instance. Experiments across multiple open test sets demonstrate URGENT-PK's superior system-level ranking performance over state-of-the-art baselines, despite its simple network architecture and limited training data.
\end{abstract}

\begin{IEEEkeywords}
speech evaluation, scoring model, Mean Opinion Score, comparison-based method
\end{IEEEkeywords}

\section{Introduction}

The Mean Opinion Score (MOS) is generally regarded as the gold standard in speech quality assessment (SQA), with wide applications in speech synthesis, speech enhancement (SE), and other areas. However, since it requires subjective evaluation by human listeners, obtaining the MOS requires substantial human resource expenditure and has relatively low efficiency. To circumvent this problem, researchers have been developing neural network (NN) based models for MOS prediction directly from the speech signal, such as DNSMOS \cite{dnsmos,dnsmos_p835}, UTMOS \cite{utmos,utmosv2} and KyotoMOS \cite{kyotomos}. Although effective in SQA for in-domain data, these NN-based methods are generally sensitive to domain mismatch and suffer from insufficient labeled data and/or increased cost to collect such data, which may lead to unreliable predictions in unseen conditions. Although numerous approaches have been proposed to address this issue, such as self-supervised learning (SSL) \cite{utmos,utmosv2,gamos,kyotomos}, ensemble learning \cite{utmos,utmosv2,kyotomos}, and introducing prior knowledge \cite{dnsmos_p835,ldnet}, the data scarcity issue remains unresolved.

Moreover, MOS labels can themselves be noisy due to the subjective nature of the listening test.
For example, when the ITU-T P.808 standard~\cite{P.808,Open-Naderi2020} is adopted, subjects are asked to rate a list of given speech samples with absolute category ratings (from 1 to 5), where the implicit rating standard can vary significantly depending on the subjects' personal preference.
This can easily lead to incomparable MOS ratings across different listening tests, making it less effective to combine multiple existing SQA datasets for training.
In contrast to absolute rating, humans are better at comparing relative speech quality, i.e., determining whether one speech sample sounds better than another.
Such a simplified design, also known as the A/B test, can provide a more consistent scoring across listening tests. 

Inspired by the above observation, in this paper, we propose a novel ranking model leveraging pairwise comparisons, named URGENT-PK (also abbreviated as `UG-PK' in this paper). `PK' is the abbreviation of the phrase `Player Kill', where two players compete with each other, and a winner is determined.
URGENT-PK is trained to predict the relative speech quality from a pair of input speeches.
Here, we associate this model with the speech enhancement (SE) task, where the model is used to rank various SE models in a strategic manner.

URGENT-PK has two components: an utterance-level pairwise model and a system-level ranking algorithm. The pairwise model takes paired speech samples from different enhancement systems processing the same noisy input, then outputs a comparative score $p$ where $p > 0.5$ indicates that the first sample has higher quality and $p <= 0.5$ indicates the opposite condition.
The pairwise learning methodology has been applied in previous work for channel selection, named MicRank~\cite{micrank}, and can also be extended to other tasks, such as speech synthesis \cite{speech_syn_review}, voice conversion \cite{voice_conv_overview}, and information retrieval RankNet~\cite{liu2009learning}.
Unlike RankNet~\cite{liu2009learning} and MicRank~\cite{micrank}, URGENT-PK explicitly models the pairwise comparison in the architecture via a comparison module, and directly estimates the winner of the pairwise input. 
As such, the system-level ranking algorithm traverses through all the binary permutations among the systems, for each pair, the algorithm then compares all the homologous enhanced speech pairs to accumulate scores for the systems. Finally, the systems are ranked according to their accumulated scores. This system-level ranking algorithm is another crucial difference from RankNet, where inference is instead performed via the model forward pass.


\begin{figure*}[t!]
\centering
\includegraphics[width=0.7\linewidth]{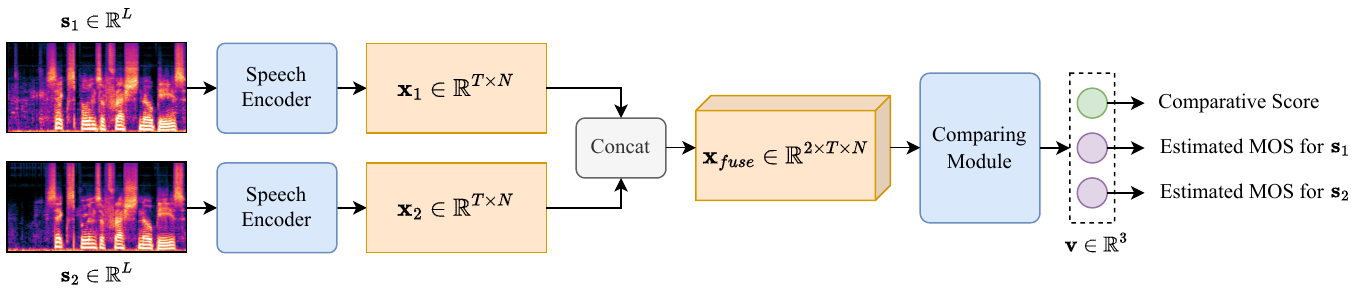}
\caption{The overall structure of the utterance-level pairwise model.}
\label{fig:URGENT-PK_model}
\end{figure*}

The URGENT-PK models are trained on the urgent24 \cite{urgent2024} dataset and tested on the urgent25 \cite{urgent2025} dataset as well as the CHiME-7 UDASE dataset \cite{chime7}. The utilized datasets' information will be detailed in Section \ref{sec:dataset}. Two widely used MOS estimation models, DNSMOS \cite{dnsmos,dnsmos_p835} and UTMOS \cite{utmos}, are chosen as baselines. Experimental results demonstrate that our proposed URGENT-PK ranking model comprehensively outperforms the state-of-the-art baseline models, meanwhile showing stronger generalization capabilities, even with a rather simple and lightweight model architecture and very limited training data.


\section{Pairwise Comparison-based Ranking Model}
\label{sec:URGENT-PK}

\subsection {Utterance-level Pairwise Model}
\label{sec:comparing_model}

The utterance-level pairwise model receives a pair of speech samples and compares their quality through neural networks. We apply a simple and intuitive Encoder-Comparing Module architecture as shown in Figure \ref{fig:URGENT-PK_model}. The speech encoder first processes the input pair of speech samples $\textbf{s}_1$ and $\textbf{s}_2$ separately, and extracts a pair of temporal embeddings $\textbf{x}_1$ and $\textbf{x}_2$, where $\textbf{s}_1,\textbf{s}_2 \in \mathbb{R}^{L}$ and $\textbf{x}_1,\textbf{x}_2 \in \mathbb{R}^{T \times N}$, $L$ denotes the speech length, $T$ and $N$ denote the temporal dimension and the feature dimension, respectively. The pair of embeddings are concatenated by an additional channel dimension to form a fused feature $\operatorname{Concat}[\textbf{x}_1, \textbf{x}_2] \in \mathbb{R}^{2 \times T \times N}$. The comparing module is then employed to predict a comparative score $\text{SCORE}_{\mathrm{cp}} \in [0,1]$ from the fused feature. To ensure that the pairwise model captures knowledge about human ear's perception of speech quality, it is designed to additionally predict the MOS of both input speech samples, denoted as $\text{MOS}_{\mathrm{pre}}^{1}$ and $ \text{MOS}_{\mathrm{pre}}^{2}$. For the sake of clarity, in this paper, we set $\text{SCORE}_{\mathrm{cp}}$, $\text{MOS}_{\mathrm{pre}}^{1}$ and $\text{MOS}_{\mathrm{pre}}^{2}$ callable to denote the usage of each output. Note that the use of such a pairwise comparing module makes the proposed approach fundamentally differeserent from RankNet~\cite{liu2009learning} and MicRank~\cite{micrank}, which enforce ranking relationships only implicitly through their loss functions rather than through dedicated architectural components.

Now, we explain the details about each module:

\subsubsection{Log-mel Spectrum Encoder}

Since the proposed model is aimed at simulating the human ear's perception, the first choice of the speech encoder in the utterance-level pairwise model is the log-mel spectrum. Unlike conventional frequency representations that operate on a linear scale, the log-mel spectrum employs the mel scale to mimic the way human perceives pitch differences, where changes in pitch are perceived more acutely at lower frequencies than at higher frequencies. The number of mel filters is set to 120 in our experiments.



\subsubsection{UTMOS-based Encoder}

UTMOS~\cite{utmos} is a state-of-the-art MOS prediction model that employs a sophisticated ensemble learning framework combining strong and weak learners to achieve superior performance. At its core, the strong learner processes raw speech waveforms through a pre-trained self-supervised learning (SSL) model to extract frame-level features, which are then passed through a BiLSTM network and linear layer to predict frame-level MOS rather than using averaged features. Furthermore, to address the variability in human ear's perception, UTMOS incorporates listener-specific embeddings concatenated with SSL features, while domain IDs help adapt to different datasets. The system further boosts accuracy through phoneme encoding, where an ASR model transcribes utterances into phoneme sequences that are processed alongside clustered reference texts using a dedicated phoneme encoder. To prevent overfitting, especially for the data-scarce out-of-domain track, UTMOS also applies carefully calibrated data augmentation techniques like pitch-shifting and speaking-rate modification. 

In this paper, we adopt UTMOS as an alternative encoder, as it was the top MOS predictor in the VoiceMOS Challenge 2022 \cite{voicemos2022} and has shown high correlations to MOS obtained from the subjective listening tests in the URGENT 2024 \cite{urgent2024} and 2025 \cite{urgent2025} Challenges, surpassing DNSMOS \cite{dnsmos} and NISQA \cite{NISQA}. In this way, we can investigate the impact of incorporating prior knowledge about human perception in the URGENT-PK model. We remove the final layer of the UTMOS model to extract the 1024-dimensional latent feature.

\subsubsection{ResNet-based Comparing Module}

To ensure that the embeddings of the two speech samples, $\textbf{x}_1$ and $\textbf{x}_2$, are sufficiently processed for accurate comparison, in this paper, we employ the modified ResNet34~\cite{resnet34,resnet_wang} architecture as the comparison module. ResNet34 is a 34-layer deep convolutional neural network (CNN) that introduces residual learning with skip connections, which has been widely utilized for speech feature processing \cite{resnet34use1,resnet34use2,cumlinDNSMOSProReducedSize2024}.

Considering the input shape $(2,T,N)$ of the fused feature as introduced in Section \ref{sec:comparing_model}, the ResNet34 begins with a convolutional layer equipped with 32 filters and kernel size of $3 \times 3$, followed by a batch normalization layer and a Rectified Linear Unit (ReLU) activation function. This initial layer is designed to capture basic features from the input data, yielding an output embedding of $(32,T,N)$.

Subsequently, the model undergoes a four-stage processing procedure, with each stage employing several residual blocks, where the number is set to $[3,4,6,3]$. In each residual block, there are two convolutional layers, each followed by a batch normalization layer and a ReLU activation function, the kernel size is set to a fixed shape of $3 \times 3$. In the first stage, the stride of all the convolutional layers is set to 1, whereas in the remaining three stages, the stride of the first convolutional layer in the first residual block is set to 2 and the rest are set to 1. This four-stage process yields an output embedding of $(256,\frac{T}{8},\frac{N}{8})$.

After that, the model performs both mean pooling and variance pooling along the temporal dimension and concatenates the outputs together by the feature dimension. This yields an output embedding of $\mathbb{R}^{64 \times N}$. In the end, a linear layer is employed to obtain the comparative score $\text{SCORE}_{\mathrm{cp}}$, as well as the estimated MOS for both inputs $\text{MOS}_{\mathrm{pre}}^1$ and $\text{MOS}_{\mathrm{pre}}^2$, yielding a final output of $\mathbb{R}^3$.

\subsection{System-level Ranking Algorithm}
\label{sec:ranking_algorithm}

In Section \ref{sec:comparing_model}, we propose an utterance-level pairwise model that accepts two homologous speeches as input, and outputs a comparative score, as well as the estimated MOS. This model is designed to find out the one with higher speech quality from a speech pair. However, in real application scenarios, the MOS listening test mainly aims at obtaining the final system-level rankings instead of utterance-wise comparisons. To this end, based on the utterance-level comparison, a system-level Enumerating-Comparing-Scoring (ECS) ranking algorithm is proposed, as shown in Algorithm \ref{alg:ranking_algorithm}.


\begin{algorithm}[htbp]
    \SetAlgoLined
    \caption{System-level ECS Ranking Algorithm}
    \label{alg:ranking_algorithm}
    \KwIn{The set of enhanced speeches: $\{\mathbf{s}_k^i\}$, where $\mathbf{s}_k^i$ denotes the enhanced speech of the $i^{th}$ noisy speech in the dataset by the $k^{th}$ system.}
    \KwOut{The accumulated score of each system: $\{p_k\}$}
    $\mathcal{P} = \{(k, w)|k, w \in \{1, 2, ..., K\}, k<w\}$ \\
    $p_k = 0, k \in \{1,2,...,K\}$ \\
    \For{ $ (k,w) \in \mathcal{P}$}{
        \tcp{compare system $k$ and system $w$}
        \For{$i \in \{1, 2, ..., M\}$}{
            $score = \text{SCORE}_{\mathrm{cp}}(\mathbf{s}_k^i,\mathbf{s}_w^i)$ \\
            \uIf{Binary Scoring}{
                \uIf{$score > 0.5$}{
                    $p_k = p_k + 1$ \tcp{system $k$ wins}
                }
                \uElse{
                    $p_w = p_w + 1$ \tcp{system $w$ wins}
                }
            }
            \uElse{  
                $p_k = p_k + score$ \\
                $p_w = p_w + (1 - score)$
            }
        }
    }
    \Return $\{p_k\}$ \\
\end{algorithm}
We assume that there are a total of $K$ systems to be ranked, each producing $M$ enhanced speech samples from the same noisy speech dataset. The enhanced speeches are denoted as $\{\mathbf{s}_k^i\}$, where $\mathbf{s}_k^i$ denotes the enhanced speech of the $i^{th}$ noisy speech in the dataset by the $k^{th}$ system, $k \in \{1, 2, ..., K\}$ and $i \in \{1, 2, ..., M\}$. The proposed ECS ranking algorithm first traverses all pairs of the $K$ systems, resulting in a total of $\frac{K \times (K-1)}{2}$ iterations. For each pair of systems $(k, w)$, the algorithm then traverses each pair of enhanced speeches $(\mathbf{s}_k^i, \mathbf{s}_w^i)$, produced by the two systems from the same $i^{th}$ noisy speech. Each speech pair is fed into the utterance-level pairwise model to generate a comparative score, based on which the ranking algorithm then assigns scores to both of the systems. This yields a total of $\frac{K \times (K-1)}{2} \times M$ iterations. Given that the comparative score ranges from $0$ to $1$, we further design two scoring strategies: a Binary Scoring (BS) strategy and a non-Binary Scoring (NBS) strategy.

\subsubsection{Binary Scoring Strategy}
\label{binary_scoring}

In the Binary Scoring (BS) strategy, for each pair of input speech samples, the algorithm assigns one point to the system offering the higher-quality speech, and zero points to the system offering the lower-quality speech. 

\subsubsection{Non-Binary Scoring Strategy}
\label{sec:non_binary_scoring}

In the Binary Scoring (BS) strategy, only the system offering the higher-quality speech receives points, where the quantifiable difference between the two speech samples is overlooked. In contrast, in the non-Binary Scoring (NBS) strategy, the algorithm assigns corresponding scores to both systems based on the comparative result, assigning $\text{SCORE}_{\mathrm{cp}}$ points to the first system and $1 - \text{SCORE}_{\mathrm{cp}}$ points to the second. 




\section{Experiment}
\label{sec:experiment}

\subsection{Datasets}
\label{sec:dataset}


In this paper, the challenge submissions in the final blind test phases of the URGENT Challenge 2024~\cite{urgent2024} and URGENT Challenge 2025~\cite{urgent2025}, as well as the CHiME-7 UDASE \cite{chime7,chime7use,chime7use2} evaluation data, are chosen for experimental validation. In URGENT Challenge 2024, 22 teams competed by enhancing a total of 1000 noisy speech samples, among which the enhancement results of 300 noisy speech samples in English contained MOS labels, forming an $\text{urgent24}_{\mathrm{en}}$ subset. In URGENT Challenge 2025, 21 teams competed by enhancing a total of 900 noisy speech samples, among which the enhancement results of 600 noisy speech samples had MOS labels. These 600 speech samples are further divided into four subsets based on the four distinct languages. In the CHiME-7 UDASE dataset, four teams competed by enhancing a total of 241 noisy speech samples, among which the enhancement results of 128 noisy speech samples contained MOS labels. For each dataset, we include the set of unprocessed noisy speeches as an additional system. Detailed information about these datasets is presented in Table \ref{tab:datasets}.

\begin{table}[htbp]
\centering
\small
\caption{Detailed information of the used datasets. Note that `len(s)/system' denotes the total lengths of enhanced speeches for each system, measured in seconds.}
\label{tab:datasets}
\begin{tabular}{c|cccc}
\toprule
dataset      & lang. & \#systems & \#utt/system & len(s)/system \\
\midrule
$\text{urgent24}_{\mathrm{en}}$     & en   & 23    & 300        & 2160          \\
\midrule
$\text{urgent25}_{\mathrm{en}}$ & en   & 22    & 150        & 921          \\
$\text{urgent25}_{\mathrm{de}}$ & de   & 22    & 150        & 1171          \\
$\text{urgent25}_{\mathrm{jp}}$ & jp   & 22    & 150        & 1010          \\
$\text{urgent25}_{\mathrm{zh}}$ & zh   & 22    & 150        & 947          \\
\midrule
CHiME-7      & en   & 5     & 128        & 606          \\
\bottomrule
\end{tabular}
\end{table}


In this paper, the models are trained on the $\text{urgent24}_{\mathrm{en}}$ dataset (8 utterances for validation and 292 utterances for training), and all other aforementioned datasets are used as test sets. In particular, we employed CHiME-7 UDASE and multiple multilingual test sets from URGENT 2025 to evaluate the model's generalization capability to unseen domains and languages, respectively. 

\subsection{Data cleaning}
\label{sec:data_cleaning}

Although MOS is often considered as the gold standard, this human-evaluated metric exhibits inherent drawbacks, including subjective biases and high variability. Firstly, different scorers can assign divergent quality evaluations to the same speech sample. Secondly, distinct perceptual sensitivities exist between different scorers when comparing speech pairs. In addition, intra-scorer inconsistencies may arise due to poor attitudes and fluctuating attentional states during the MOS test. Although the impact is negligible when comparing speech pairs with significant quality discrepancies, they can influence the comparison between rather similar quality speech pairs, thus confusing the pairwise model in the training stage.
To mitigate this influence induced by the aforementioned drawbacks, following prior works \cite{micrank}, we implement a data cleaning scheme that ignores, during training, speech pairs with close MOS values, which can be considered to have indistinguishable perceptual quality.

Specifically, we set a `MOS difference threshold', $\delta$, to filter the speech pairs in the dataset. For each pair of speech samples, only when the MOS difference exceeded the threshold $\delta$ are they included in the training data. This filtering strategy can, to some extent, improve the quality of the training data, but concurrently results in the loss of data volume. In this paper, we set the score difference threshold to $\delta = 0.3$, aiming to reach a tradeoff between data quality and quantity. Ablation studies in Section \ref{sec:exp} demonstrate the rationality of this threshold setting.

\subsection{Training Details}

\subsubsection{Training Objective}

Multi-task learning is used in this paper, where the pairwise model is trained not only to compare the two input speech samples, but also to predict their MOS. In this way, the model is guided to capture knowledge about the human ears' perception of speech quality. We apply the binary cross-entropy (BCE) loss on the comparative score and the mean square error (MSE) loss on the estimated MOS. The loss function can be formulated as follows:
\begin{align}
    \label{eq:loss}
    \mathcal{L}_{\mathrm{cp}} &= \operatorname{BCE}(\text{SCORE}_{\mathrm{cp}}, \operatorname{int}(\mathrm{MOS}_1 > \mathrm{MOS}_2)), \\
    \mathcal{L}_{\mathrm{sc}} &= \operatorname{MSE}([\text{MOS}_{\mathrm{pre}}^1,\text{MOS}_{\mathrm{pre}}^2], [\mathrm{MOS}_1,\mathrm{MOS}_2]), \\
    \mathcal{L} &= \alpha \times \mathcal{L}_{\mathrm{cp}} + \beta \times \mathcal{L}_{\mathrm{sc}}.
\end{align}
where $\mathrm{MOS}_1,\mathrm{MOS}_2 \in [1.0, 5.0]$ denote the averaged MOS scores annotated by 8 listeners for first and second speech, respectively. The $\alpha$ and $\beta$ is set to 0.5 and 0.5 in the experiments.

\subsubsection{Other Training Details}

For the UTMOS encoder-based URGENT-PK models, the batch size is set to 4, the initial learning rate is set to $1.0 \times 10^{-5}$ with weight decay of $1.0 \times 10^{-6}$, and the models were trained for 15 epochs. For the log-mel spectrum-based URGENT-PK models, the batch size is set to 12, the initial learning rate is set to $1.0 \times 10^{-4}$ with a weight decay of $1.0 \times 10^{-6}$, and the models were trained for 30 epochs.
These parameters were determined by tuning each system on the validation set. 

\subsection{Evaluation Metrics}

Following previous works \cite{utmos,utmosv2}, in this paper, three correlations are chosen as the evaluation metrics: the Linear Correlation Coefficient (LCC), the Spearman Rank Correlation Coefficient \cite{spearman} (SRCC), and the Kendall Rank Correlation Coefficient \cite{KRCC} (KRCC).
LCC assumes normality and linearity, while both SRCC and KRCC serve as non-parametric alternatives requiring fewer distributional assumptions.
For all metrics, the higher is the better.
In the experiments, system-level correlations are calculated between the accumulated scores by the ECS ranking algorithm and the oracle average MOS.

\subsection{Validation Strategy}

Throughout the training stage, we save the best $9$ model checkpoints based on the objective functions on the validation set. For each of the saved checkpoints, we perform the system-level ECS ranking algorithm on the validation set and calculate the three above-mentioned correlation, LCC, SRCC and KRCC. We sum up the three correlations and select the checkpoint with the highest one for testing. 

\subsection{Experimental Results}

\subsubsection{Comparison with Baselines}

In this paper, two widely-used speech quality evaluation model, DNSMOS \cite{dnsmos} (overall MOS) and UTMOS \cite{utmos}, are chosen as the baselines. To make a fair comparison, the UTMOS model is also fine-tuned on the training set $\text{urgent24}_{\mathrm{en}}$. Furthermore, since the proposed ECS ranking algorithm can also combine with other MOS prediction models by comparing the estimated score, we also build a $\text{UTMOS}_{pk}$ system, which performs the ECS ranking algorithm and the BS strategy, comparing the UTMOS scores of the two speech samples each time.

\begin{table*}[t!]
\centering
\small
\setlength{\tabcolsep}{5pt}
\caption{Generalization capability of the proposed URGENT-PK model on multiple out-domain test sets comparing with the baselines, measured by correlations between the models' output score and the oracle MOS. Models are trained on $\text{urgent24}_{\mathrm{en}}$.}
\label{tab:Generalization}
\begin{threeparttable}
\begin{tabular}{c|c|ccc|ccc|ccc|ccc}
\toprule
\multicolumn{2}{c|}{dataset} & \multicolumn{3}{c|}{$\text{urgent25}_{\mathrm{zh}}$} & \multicolumn{3}{c|}{$\text{urgent25}_{\mathrm{jp}}$} & \multicolumn{3}{c|}{$\text{urgent25}_{\mathrm{de}}$} & \multicolumn{3}{c}{CHiME-7 UDASE\tnote{2}} \\
\midrule
Model                             & Strategy & KRCC & SRCC & LCC & KRCC & SRCC & LCC & KRCC & SRCC & LCC & KRCC & SRCC      & LCC       \\
\midrule
DNSMOS & / & 0.472 & 0.606 & 0.831 & 0.446 & 0.598 & 0.816 & 0.489 & 0.612 & 0.795 & 0.000 & -0.100 & -0.140 \\
UTMOS                              & /        & 0.524                    & 0.666                    & 0.665                   & 0.602                    & 0.775                    & 0.822                   & 0.671                    & 0.857                    & 0.854                   & 0.200                    & 0.200                         & 0.267                         \\
$\text{UTMOS}_{ft}$                          & /        &  0.655    &   0.824   &  0.899       &   0.686     &  0.842   &   0.914 &      0.760   & 0.891    &   0.920   &  0.067     &  0.371     &   0.316 \\
$\text{UTMOS}_{pk}$                          & /        & 0.567 & 0.718 & 0.734 & 0.576 & 0.743 & 0.773 & 0.671 & 0.829 & 0.821 & 0.000 & -0.100 & 0.100     \\
\midrule
                                   & BS       & 0.619                    & 0.815                    & 0.895                   & 0.703                    & 0.857                    & 0.920                   & 0.749                    & 0.905                    & 0.929                   & 0.200                    & 0.300                         & 0.289                         \\
\multirow{-2}{*}{$\text{UG-PK}_{\text{mel}}$}       & NBS      & 0.593                    & 0.791                    & 0.896                   & 0.723                    & 0.870                    & \textbf{0.921}                   & 0.740                    & 0.902                    & 0.932                   & 0.200                    & 0.300                         & 0.283                         \\
\midrule
                                   & BS       & 0.628                    & 0.778                    & 0.825                   & 0.703                    & 0.852                    & 0.849                   & 0.810                    & 0.944                    & 0.905                   & \textbf{0.400}                    & \textbf{0.500}                         & 0.402                         \\
\multirow{-2}{*}{$\text{UG-PK}_{\text{UTMOS}(fix)}$}     & NBS      & 0.654                    & 0.797                    & 0.850                   & 0.714                    & 0.875                    & 0.878                   & 0.810                    & 0.940                    & 0.922                   & \textbf{0.400}                    & \textbf{0.500}                         & 0.382                         \\
\midrule
                                   & BS       & 0.706                    & 0.888                    & 0.929                   & 0.706                    & 0.880                    & 0.891                   & \textbf{0.835}                    & \textbf{0.951}                    & 0.933                   & \textbf{0.400}                    & \textbf{0.500}                         & \textbf{0.489}                         \\
\multirow{-2}{*}{$\text{UG-PK}_{\text{UTMOS}(ft)}$} & NBS      & \textbf{0.723}                    & \textbf{0.893}                    & \textbf{0.930}                   & \textbf{0.732}                    & \textbf{0.896}                    & 0.895                   & \textbf{0.835}                    & \textbf{0.951}                    & \textbf{0.934}                   & \textbf{0.400}                    & \textbf{0.500}                         & 0.485  \\
\bottomrule
\end{tabular}
\begin{tablenotes}
    \footnotesize
    \item[2] Note that the calculated KRCC and SRCC exhibit a coarse resolution because there are only 5 systems in CHiME-7 UDASE.
\end{tablenotes} 
\end{threeparttable}
\end{table*}

The models are trained on the $\text{urgent24}_{\mathrm{en}}$ dataset. First, we compare our proposed model with the baselines on the $\text{urgent25}_{\mathrm{en}}$ dataset. As shown in Table \ref{tab:urgent_en}, our proposed pairwise comparison-based URGENT-PK models consistently outperform the baselines. Even the $\text{URGENT-PK}_{\mathrm{mel}}$ system trained from scratch with limited data shows a better overall performance than the fine-tuned UTMOS. Note that the latter possesses a certainly more powerful speech encoder pretrained on thousands of hours of data. In addition, as for the URGENT-PK models with different encoders, the mel spectrum-based model shows comparable performance with the UTMOS-based model, where the former performs better on KRCC and SRCC and the latter performs better on LCC. These experimental results fully demonstrate the superiority of our proposed comparison-based ranking approach, where considerable performance is achieved even with a simple model architecture and limited training data. Regarding the upper performance of the URGENT-PK model, the model with fine-tuned UTMOS achieves the best performance. This is expected as this system simultaneously possesses the most complex model architecture and has been trained with the most training data.

\begin{table}[htbp]
\centering
\small
\caption{Experimental results of the proposed URGENT-PK model comparing with the baselines, measured by correlations between the models' output score and the oracle MOS. Models are trained on $\text{urgent24}_{\mathrm{en}}$ and tested on $\text{urgent25}_{\mathrm{en}}$.}
\label{tab:urgent_en}
\begin{tabular}{c|c|ccc}
\toprule
Model                             & Strategy & KRCC  & SRCC  & LCC   \\
\midrule
DNSMOS                             & /        & 0.602 & 0.713 & 0.847 \\
UTMOS                              & /        & 0.835 & 0.951 & 0.894 \\
$\text{UTMOS}_{ft}$                          & /        &   0.814 &  0.944    &   0.939    \\
$\text{UTMOS}_{pk}$                           & /        & 0.766 & 0.928 & 0.842 \\
\midrule
\multirow{2}{*}{$\text{UG-PK}_{\text{mel}}$}        & BS       & 0.835 & 0.955 & 0.930 \\
                                   & NBS      & 0.853 & 0.960 & 0.935 \\
\midrule
\multirow{2}{*}{$\text{UG-PK}_{\text{UTMOS}(fix)}$} & BS       & 0.827 & 0.954 & 0.938 \\
                                   & NBS      & 0.853 & 0.957 & \textbf{0.965} \\
\midrule
\multirow{2}{*}{$\text{UG-PK}_{\text{UTMOS}(ft)}$}  & BS       & \textbf{0.879} & 0.971 & 0.955 \\
                                   & NBS      & \textbf{0.879} & \textbf{0.972} & 0.959 \\
\bottomrule
\end{tabular}
\end{table}

Subsequently, more out-of-domain test sets are included to evaluate the generalization capability of our proposed URGENT-PK model, including three multilingual subsets from the URGENT Challenge 2025, as well as the CHiME-7 UDASE evaluation dataset. Experimental results show that our proposed URGENT-PK model still generally outperforms the baselines, where the only exception is that the mel spectrogram-based URGENT-PK model performs slightly worse than the fine-tuned UTMOS on $\text{URGENT25}_{\mathrm{zh}}$. As for different URGENT-PK models, a similar performance trend is observed, where the mel spectrum-based model and the UTMOS-based model performs comparably and the model with fine-tuned UTMOS yields the best performance.

\begin{table}[htbp]
\centering
\small
\setlength{\tabcolsep}{3pt}
\caption{The A/B Test of the utterance-level pairwise model. The comparative accuracy of the pairwise models as well as the MOS comparison are measured referring to the subjective comparative label. Models are trained on $\text{urgent24}_{\mathrm{en}}$ and tested on $\text{urgent25}_{\mathrm{en}}$, and abbreviated by the speech encoder.}
\label{tab:compring_model}
\setlength{\tabcolsep}{2pt}
\begin{tabular}{c|ccccc|c}
\toprule
UG-PK Models & {[}0,0.4) & {[}0.4,0.8) & {[}0.8,1.2) & {[}1.2,1.6) & {[}1.6,2) & Avg \\
\midrule
$\text{MOS}_{\mathrm{cp}}$ & 0.70 & 0.80 & 0.95 & 1.00 & 1.00 & 0.89 \\
\midrule
mel & 0.75 & 0.55 & 0.85 & 0.85 & 0.95 & 0.79 \\
$\text{UTMOS}_{(fix)}$ & 0.85 & 0.70 & 0.80 & 0.95 & 1.00 & 0.86 \\
$\text{UTMOS}_{(ft)}$  & 0.80 & 0.65 & 0.90 & 0.95 & 1.00 & 0.86 \\
\bottomrule
\end{tabular}
\end{table}

\subsubsection{Subjective A/B Test of the Utterance-level Pairwise Model}

The performance of the utterance-level pairwise model is then investigated by a subjective A/B Test on speech quality comparison. Specifically, test speech pairs are divided into several groups based on a different range of MOS differences, where 20 speech pairs are randomly selected for each group. Subjective A/B test is conducted on these selected pairs, where 5 human participants listened to and labeled each pair, selecting the higher-quality speech. The comparison accuracy of the pairwise model, as well as the MOS comparison, is measured with reference to the subjective label. Note that speech pairs with MOS differences larger than 2 are ignored as the comparative accuracy is consistently $1.00$.

The experimental results are shown in Table \ref{tab:compring_model}, from which we can summarize several key findings as follows: Firstly, larger MOS difference generally brings better performance for the pairwise model, except for the group with MOS difference in the range of $[0.4,0.8)$. This is acceptable considering that the limited quantity of test data may cause performance fluctuations. Secondly, the accuracy of MOS comparison becomes pretty low when the MOS difference is quite small, even lower than the pairwise model in the range of $[0.0,0.4)$. This finding preliminarily signifies the necessity of training data cleaning, which will be further investigated in Section \ref{sec:ablation_data_cleaning}. Lastly, in general, our proposed pairwise models achieved considerable performance in the subjective A/B test, with their comparative accuracy reaching a level comparable to that of the MOS comparison, or even outperforming it in the range of $[0.0,0.4)$.

\subsubsection{Ablation Study on the Predicted MOS}
\label{sec:ablation_MOS}

The utterance-level pairwise model not only generates a comparative score, but also estimates the MOS of the input speech samples. Although the predicted MOS are primarily designed for multi-task learning, they enable the model to directly serve as a speech quality evaluation model. In order to verify whether the pairwise model has learned about human ear's perception, we include an ablation study to assess the performance of MOS prediction. Two strategies are designed to transform the pairwise model into a speech quality assessment (SQA) model: the Replication Strategy and the Noisy-Speech Strategy.

In the Replication Strategy, the enhanced speech sample $\mathbf{s}$ is replicated and fed into the pairwise model to get the predicted MOS, the average of both the predicted MOS is used as the final evaluation score, which can be formulated as $\text{MOS}_{\mathrm{pre}}={(\text{MOS}_{\mathrm{pre}}^1(\mathbf{s},\mathbf{s})+\text{MOS}_{\mathrm{pre}}^2(\mathbf{s},\mathbf{s}))}/{2}$.

In the Noisy-Speech Strategy, the enhanced speech sample $\mathbf{s}$ and its corresponding unprocessed noisy speech $\mathbf{n}$ are fed into the pairwise model. Their positions are swapped to go through the model twice, thus two predicted MOS are obtained and the average is used as final evaluation score, which can be formulated as $\text{MOS}_{\mathrm{pre}}={(\text{MOS}_{\mathrm{pre}}^1(\mathbf{s},\mathbf{n})+\text{MOS}_{\mathrm{pre}}^2(\mathbf{n},\mathbf{s}))}/{2}$, 

\begin{table}[htbp]
\centering
\small
\caption{Ablation study of the predicted MOS, measured by the correlations between the predicted MOS and the oracle MOS. Models are trained on $\text{urgent24}_{\mathrm{en}}$ and tested on $\text{urgent25}_{\mathrm{en}}$}
\label{tab:use_p_score}
\begin{tabular}{c|c|ccc}
\toprule
Model                 & Strategy  & KRCC  & SRCC  & LCC   \\
\midrule
DNSMOS                             & /        & 0.602 & 0.713 & 0.847 \\
UTMOS                              & /        & 0.835 & 0.951 & 0.894 \\
\midrule
\multirow{2}{*}{$\text{UG-PK}_{\text{mel}}$}   
                              & Replication & 0.706 & 0.886 & 0.928 \\
                              & Noisy-Speech & 0.766 & 0.914 & 0.937 \\
\midrule
\multirow{2}{*}{$\text{UG-PK}_{\text{UTMOS}(fix)}$} 
                              & Replication & 0.844 & 0.950 & 0.950 \\
                              & Noisy-Speech & 0.861 & 0.962 & 0.960 \\
\midrule
\multirow{2}{*}{$\text{UG-PK}_{\text{UTMOS}(ft)}$} 
                              & Replication & \textbf{0.905} & \textbf{0.974} & \textbf{0.987} \\
                              & Noisy-Speech & 0.887 & 0.971 & 0.966 \\
\bottomrule
\end{tabular}
\end{table}

Experimental results are presented in Table \ref{tab:use_p_score}, where several conclusions can be drawn as follows: Firstly, considerable performance is achieved when utilizing the utterance-level pairwise model as an utterance evaluation model, which suggests that the pairwise model has indeed captured knowledge about human ear's perception. Secondly, the UTMOS-based pairwise model significantly outperforms the mel spectrum-based model. This is predictable since UTMOS is specifically designed for SQA and has already learned substantial knowledge about SQA. However, the mel spectrum-based pairwise model still comprehensively outperforms DNSMOS, and outperforms UTMOS in terms of LCC, demonstrating the superiority of the proposed pairwise model despite its limited training data.

\subsubsection{Ablation Study on the Data Cleaning}
\label{sec:ablation_data_cleaning}

\begin{figure}[htbp]
\centering
\includegraphics[width=0.9\linewidth]{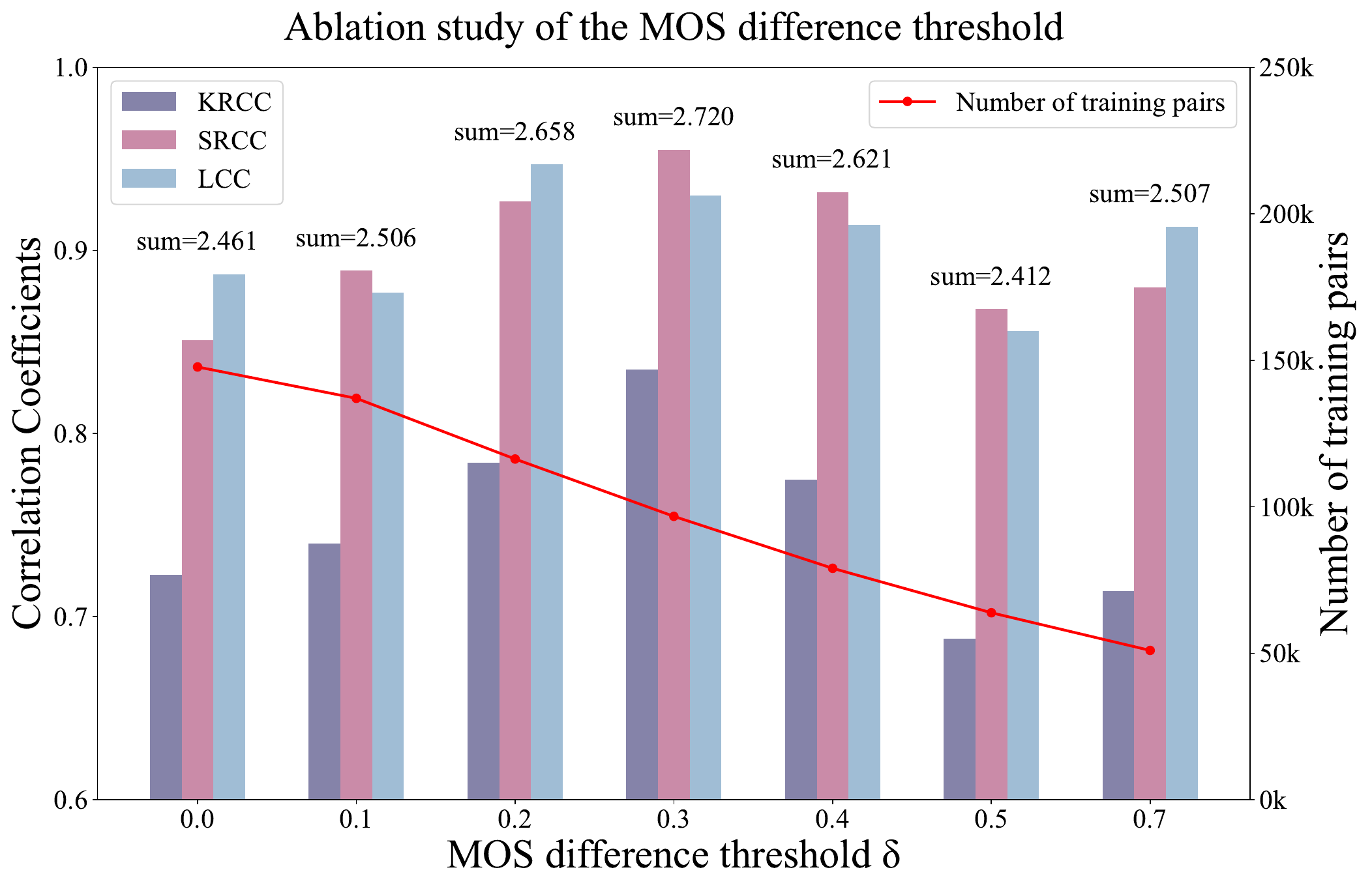}
\caption{Ablation study of the MOS difference threshold $\delta$ in data cleaning. Models are trained on $\text{urgent24}_{\mathrm{en}}$ and tested on $\text{urgent25}_{\mathrm{en}}$}
\label{fig:ablation_study}
\end{figure}

Finally, the ablation study on the data cleaning introduced in \ref{sec:data_cleaning}, is conducted by setting various MOS difference threshold, $\delta$, and testing the model's performance. Figure \ref{fig:ablation_study} shows the experimental results, where each $\delta$ yields a different amount of training data. The $\text{URGENT-PK}_{\mathrm{mel}}$ system and the Binary Scoring strategy is employed in the ablation study.

Based on the bar graph presented in Figure \ref{fig:ablation_study}, we can draw the following conclusions: First, when the difference threshold $\delta$ is rather small, gradually increasing $\delta$ consistently produces better performance, since a larger $\delta$ value removes more confusing training speech pairs from the dataset without losing too much training data. However, after $\delta$ reaches $0.3$, which is the experimental setting in this paper, continuing to increase $\delta$ will deteriorate the model’s performance. This is because a larger MOS difference implies a more evident difference in quality, indicating that the speech pairs are no more confusing. At this point, increasing $\delta$ only results in the loss of training data, thus reducing the capability and robustness of the pairwise model.

\label{sec:exp}

\section{Conclusion}
\label{sec:conclusion}

In this paper, we propose a pairwise comparison-based URGENT-PK ranking model designed for speech enhancement competitions. URGENT-PK consists of an utterance-level pairwise model and a system-level ranking algorithm. The pairwise model takes two homologous speech samples as input and generates a comparative score. The proposed Enumerating-Comparing-Scoring ranking algorithm traverses all the system pairs and accumulates scores for each system. Extensive experiments demonstrate that URGENT-PK surpasses state-of-the-art NN-based SQA baselines when facing the real data of speech enhancement competitions, even with a simple network architecture and limited training data. In future work, we aim to improve the model's performance further, as well as to extend the proposed method to a broader range of tasks.

\clearpage
\bibliography{reference}

\end{document}